\documentclass{article}

\usepackage[preprint,nonatbib]{neurips_mlsb_2020}
\usepackage[square,numbers]{natbib}
\usepackage[utf8]{inputenc} 
\usepackage[T1]{fontenc}    
\usepackage{hyperref}       
\usepackage{url}            
\usepackage{booktabs}       
\usepackage{amsfonts}       
\usepackage{nicefrac}       
\usepackage{microtype}      
\usepackage{amsmath} 
\usepackage{graphicx}
\usepackage[font=small,labelfont=bf]{caption}

\title{Cryo-ZSSR: multiple-image super-resolution based on deep internal learning}

    \author{Qinwen Huang$^1$ , Ye Zhou$^1$ , Xiaochen Du$^{1,2}$ , Reed Chen$^3$ , Jianyou Wang$^1$, 
    \\\textbf{Cynthia Rudin}$^{1,4,5}$, \textbf{Alberto Bartesaghi}$^{1,5,6}$ \\
    \\
    $^1$ Department of Computer Science, Duke University\\
    $^2$ Department of Chemistry, Duke University\\
    $^3$ Department of Biomedical Engineering, Duke University\\
    $^4$ Department of Statistical Science, Duke University\\
    $^5$ Department of Electrical and Computer Engineering, Duke University\\
    $^6$ Department of Biochemistry, Duke University School of Medicine\\
    \\
     \texttt{\{qinwen.huang, ye.zhoue678, xiaochen.du, jianyou.wang\}@duke.edu}\\
    \texttt{\{cynthia, alberto\}@cs.duke.edu}}

\begin{document}

\maketitle

\begin{abstract}
  Single-particle cryo-electron microscopy (cryo-EM) is an emerging imaging modality capable of visualizing proteins and macro-molecular complexes at near-atomic resolution. The low electron-doses used to prevent sample radiation damage, result in images where the power of the noise is 100 times greater than the power of the signal. To overcome the low-SNRs, hundreds of thousands of particle projections acquired over several days of data collection are averaged in 3D to determine the structure of interest. Meanwhile, recent image super-resolution (SR) techniques based on neural networks have shown state of the art performance on natural images. Building on these advances, we present a multiple-image SR algorithm based on deep internal learning designed specifically to work under low-SNR conditions. Our approach leverages the internal image statistics of cryo-EM movies and does not require training on ground-truth data. When applied to a single-particle dataset of apoferritin, we show that the resolution of 3D structures obtained from SR micrographs can surpass the limits imposed by the imaging system. Our results indicate that the combination of low magnification imaging with image SR has the potential to accelerate cryo-EM data collection without sacrificing resolution.
\end{abstract}

\section{Introduction}
\label{intro}

Single-particle cryo-EM is a powerful imaging modality used to determine the three-dimensional structure of proteins and macro-molecular complexes at near-atomic resolution \cite{Bartesaghi2015,Bartesaghi2018,Bendory2020,Singer2020}. By combining hundreds of thousands of noisy projection images of identical copies of the molecule of interest taken from different orientations, 3D reconstructions can be obtained where molecular level interactions can be visualized. While the signal contributed by each individual projection is extremely weak, averaging the contribution from many particles allows to overcome the extremely low signal-to-noise ratios (SNR). Acquiring such large datasets is time consuming and can take several days to complete, becoming a bottleneck in the structure determination pipeline. One strategy to improve the throughput of data collection is to increase the size of the field of view by acquiring images at lower magnification. For example, doubling the pixel size will increase the image area 4-fold resulting in four times as many particles per exposure. While this strategy will limit the attainable resolution due to the coarser spatial sampling, it does not mean a permanent loss of high-frequency information because images are collected in movie-mode and random shifts exist between the acquired frames. SR techniques that enable recovery of high-frequency information from aliased low resolution signals could become a natural solution to this problem.

Many super-resolution (SR) algorithms based on machine learning have been proposed that achieve state of the art (SotA) performance on natural images. Here, we set out to explore whether these strategies can be extended to work on low-SNR images such as the low-dose projections obtained in single-particle cryo-EM. We propose to use a multiple-image SR algorithm that utilizes deep internal learning as initially presented in the zero-shot super-resolution (ZSSR) framework \cite{shocher2017zeroshot}. Learning based on internal statistics has shown promising results for cryo-EM image denoising \cite{krull2019noise2void,denoiseconv}. As multiple frames are acquired from each field of view and each exposure area contains hundreds of projections of the same macro-molecule, data repetition occurs naturally in single-particle cryo-EM. Our algorithm exploits cross scale internal data repetition in noisy movies obtained from frozen hydrated protein samples imaged under an electron microscope. We train a movie-specific neural network that takes in multiple frames from each low-resolution (LR) movie and reconstructs a single 2X SR image per exposure area. These SR images are then fed into the standard cryo-EM data processing pipeline and used to generate a 3D reconstruction of the protein of interest, {\bf Figure 1}. Our SR algorithm is self-supervised and does not require training on ground-truth data. We evaluate the performance of cryo-ZSSR on a real dataset of apoferritin and show that the SR images can produce higher resolution reconstructions compared to the LR data. Used in combination with low-magnification imaging, our approach can be used to accelerate data collection while still producing high-quality 3D reconstructions.
\begin{figure}
  \centering
  \includegraphics[width=13.5cm, height=9.8cm]{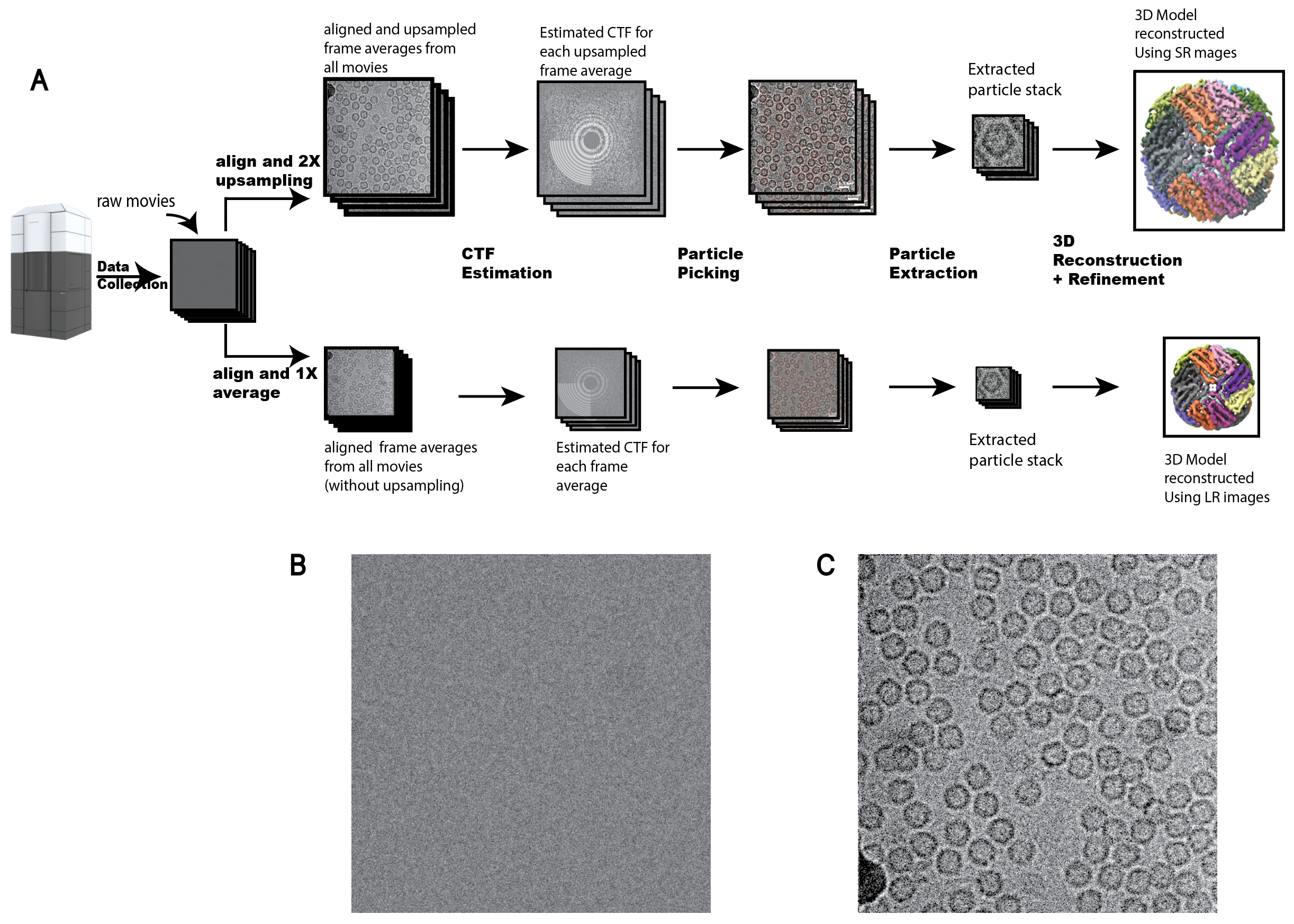}
  \caption{\underline{Cryo-ZSSR single-particle structure determination pipeline}. {\bf A. } Cryo-EM movies are collected using a large pixel size and subsequently up-sampled by a factor of 2 using our self-supervised cryo-Zero Shot Super Resolution (cryo-ZSSR) approach (top branch). Super-resolved micrographs are then fed into the standard single-particle reconstruction workflow producing three-dimensional structures at resolutions surpassing the Nyquist rate. The 2X up-sampling factor effectively results in a 4x speedup in the rate of data acquisition allowing the collection of four times more particles in the same amount of time. Standard cryo-EM structure determination pipeline without super-resolution (bottom branch). {\bf B. } Example of a single raw frame from a movie of apoferritin from EMPIAR-10146 collected at 2 $e^-/${\AA}$^2$. {\bf C. } Average of 50 frames corresponding to a total dose of 100 $e^-/${\AA}$^2$. 
}
\end{figure}
\section{Related Work}
There has been a tremendous amount of work on image SR in the past couple of decades, both for single image SR (SISR) and for multiple image SR (MISR). Traditional SISR algorithms can be divided into three categories: interpolation-based methods (e.g. bilinear, Lanczos kernels, etc.), reconstruction-based methods, and example based methods. Interpolation-based algorithms are straightforward and fast but suffer from limited accuracy. Reconstruction-based methods make use of prior knowledge about images by restricting the possible solution space to generate high-quality images. However, these methods are usually time-consuming and their performance degrades rapidly as the up-sampling factors increase. Example-based methods usually leverage machine learning to analyze relationships between the LR and its corresponding high-resolution (HR) counterparts from training examples. More recently, deep learning based SISR algorithms built upon example-based learning, have received wide attention and demonstrated great superiority compared to more traditional approaches. Generative adversarial network (GAN) \cite{goodfellow2014generative} based deep learning approaches, such as Super Resolution Generative Adversarial Network (SRGAN) \cite{ledig2016photorealistic}, and Photo Sampling via Latent Space Exploration (PULSE) \cite{menon2020pulse} are able to perform extremely well on certain natural and facial images. However, most of these networks are trained in a supervised manner and require knowledge of ground truth images. Results reconstructed using GANs, even though visually appealing, tend to generate information that does not exist in the actual HR pictures. In addition, the formation of training datasets, specifically the LR images, are usually generated using predetermined ideal processes (e.g. bicubic downsampling, gaussian blurring, etc.). In reality, LR images rarely follow this model, resulting in poor performance of previously mentioned SotA methods. To overcome this limitation, ZSSR was proposed by Shocher et al. \cite{shocher2017zeroshot}. This method does not rely on prior training. Instead, it exploits the internal recurrence of information inside a single image and trains an image-specific CNN on examples extracted solely from the input image itself. ZSSR is able to achieve and outperform SotA methods on LR images generated under non-ideal down-sampling models. \par
Turning to the problem of MISR, which involves the extraction of information from many LR observations of the same scene to reconstruct HR images, the earliest method developed by Tsai and Huang \cite{tsai} used a frequency domain technique to improve the spatial resolution of images by combining multiple LR images using sub-pixel displacements. Later on, other spatial domain MISR methods were proposed. These methods include non-uniform interpolation such as adaptive kernel regression \cite{kernelTakeda}, Bayesian modeling algorithms \cite{bayesian}, and projection onto convex sets (POCS) \cite{pocs}. Most of these SR methods assume a priori knowledge about the motion model, blur kernel and noise level. However, there are many cases where the actual image degradation process is unknown. \par
For this reason, many blind SR image reconstruction methods were developed. These methods usually involve two steps: 1) motion estimation for LR images, followed by 2) simultaneous estimation of both the HR image and the blurring function. Since separating image registration and HR estimation tends to produce sub-optimal results, some researchers have developed methods that jointly estimate motion parameters and the HR reconstruction \cite{joint}. Recently, similar to SISR problems, deep learning based methods have been proposed to simultaneously solve video SR and MISR problems. Most of the existing work is focused on video SR, such as frame recurrent SR \cite{FRVSR} which utilizes previous inferred HR frames to super-resolve subsequent frames in an end-to-end trainable framework that incorporates both frame registration and HR estimation. Most recently, a challenge was set by the European Space Agency (ESA) to super-resolve multi-temporal PROBA-V satellite imagery. In this context, multiple MISR CNN-based algorithms were proposed, including HighRes-net \cite{deudon2020highresnet} which uses an end-to-end mechanism that learns co-registration, fusion, upsampling; and residual attention model (RAMS), that utilizes 3D convolutions to exploit spatial and temporal relationships across images for HR reconstruction \cite{salvetti2020multi}. \par
Despite all the previous work of SR on natural and satellite images, little work has been done on MISR methods in the context of cryo-EM. Preliminary work done by Chen et al. \cite{CHENsr} demonstrated that MISR reconstruction surpassing the Nyquist frequency is possible by using a noiseless synthetic dataset and without considering the modulation effects of the contrast transfer function (CTF) (point spread function in spatial domain). Real micrographs acquired with an electron microscope, however, inevitably suffer from low-SNR due to the small doses used during imaging and are modulated by the CTF. Meanwhile, deep learning techniques have been applied to cryo-EM imaging in a variety of other contexts, including particle picking \cite{deeppicker,cryolo,topaz,deepconcensus}, automated micrograph and class selection \cite{micrographcleaner}, CNNs for segmentation of cryo-electron tomograms \cite{emansegment}, map denoising and local resolution estimation \cite{Tegunov2020,deepres,deepresolution}, and more recently the study of conformational heterogeneity during 3D reconstruction \cite{cryodrgn}. Inspired by recent success of deep learning MISR approaches, we aim to tackle the problem of MISR using cryo-EM images by jointly registering LR images and reconstructing SR images, all within an end-to-end trainable network based on the ZSSR framework \cite{shocher2017zeroshot}.

\section{Method}
\begin{figure}
  \centering
  \includegraphics[width=12cm,height=10cm]{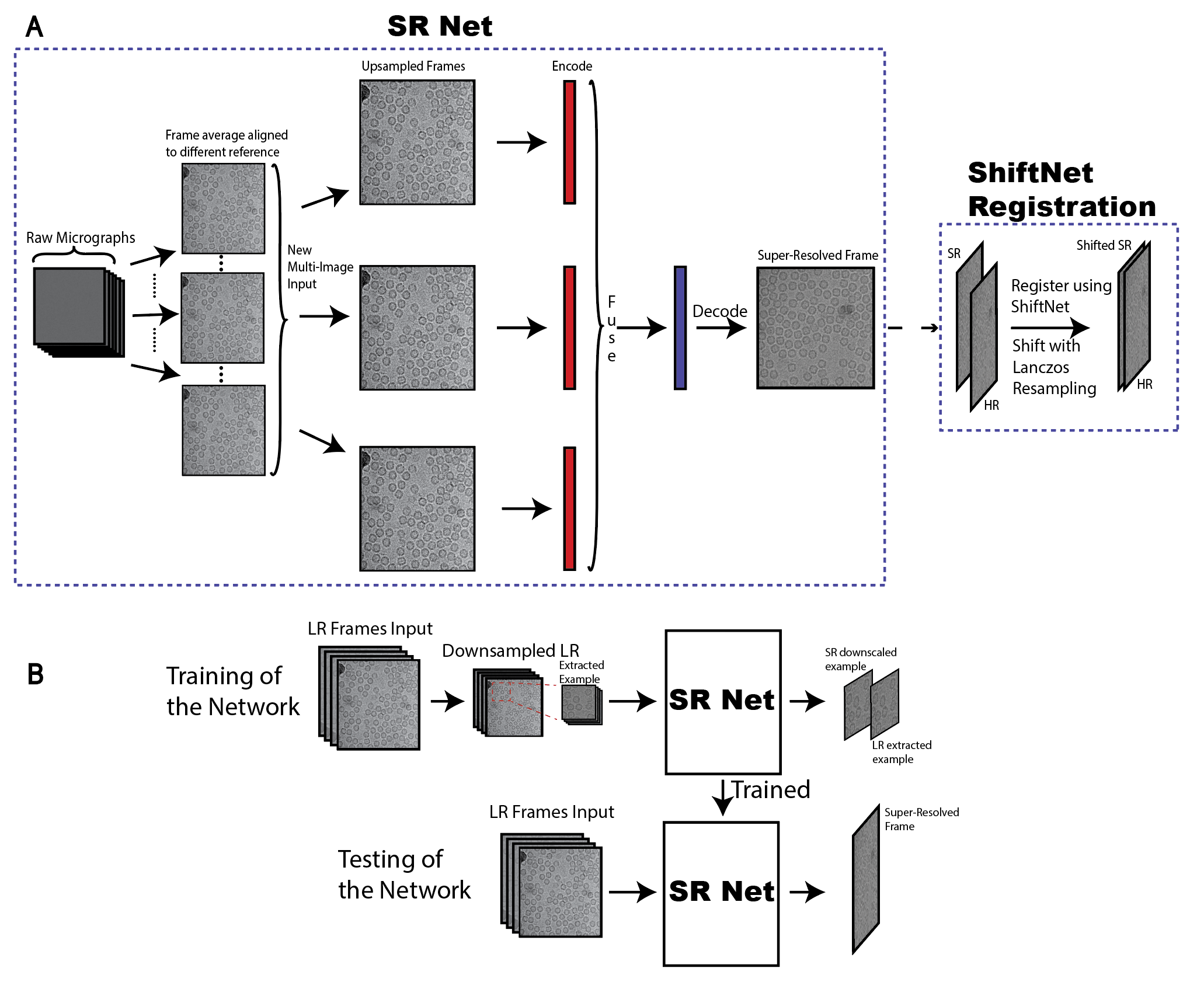}
  \caption{\underline{Overall architecture of Cryo-ZSSR.} {\bf A}. \textit{Left: }The overall framework of HighResNet based SR Net. Raw micrographs are aligned to different reference frames. A total of 16 frame averages are obtained. These frame averages now serve as LR inputs to the network. The LR inputs are first up-sampled to the size of the desired SR output (scale factor = 2). The up-sampled LR inputs then go through encode, fusion and decode stages. During the encoding stage, the up-sampled LR frames are paired with a reference frame (median of all frames) and encoded into the latent representation. The latent encodings are then fused recursively into a global encoding. The SR output is reconstructed by decoding the encoded representation. No upsampling occurs during the encoding or decoding stages. \textit{Right: }The output SR is registered with the actual HR using ShiftNet and loss after registration is calculated. {\bf B}. ZSSR based training of SR Net in (A).
}
\end{figure}
MISR aims at recovering an HR image $I^{HR}$ from a set of $K$ LR images $I^{LR}_{i}, i\in[1,...,K]$ of the same scene acquired during a certain time window. Typically, a LR image $I^{LR}_{i}$ is related to the HR image $I^{HR}$ through motion shift, blurring, down-sampling and noise corruption. In the context of cryo-EM, motion shift mainly comes from beam-induced motion that occurs during data acquisition, blurring is modeled by the CTF, and severe noise corruption is a result of low electron dosage during imaging. It is worth noting that the CTF, unlike other commonly used blurring kernels such as Gaussian, has multiple zero crossings, which means information at certain frequencies is completely lost, making direct inversion impossible. Denoising techniques applied at single image-level are also prone to the removal of high frequency information along with the actual noise. Therefore, in the standard cryo-EM data processing pipeline, CTF inversion and denoising is not applied until the final step of 3D reconstruction. In order to cope with the special characteristics of cryo-EM images described above, unlike most standard SR algorithms, instead of recovering an HR image $\Tilde{I}^{HR}$ that is free of blurring and noise from LR images, we aim to generate $I^{SR}$ subject to the modulation by the CTF and noise using our proposed methodology $G$:
\begin{equation}
\begin{split}
    I^{SR} & = G(I^{LR}_{1,..K}) \\
    I^{SR} & = H*\Tilde{I}^{HR} + \sigma
\end{split}
\end{equation}
where $\Tilde{I}^{HR}$ is the ground truth HR image without CTF modulation and free of noise (which can be viewed as the actual sample), $H$ is the same CTF that modulates both LR and HR images at different frequency ranges, and $\sigma$ represents the zero-mean Gaussian noise. Since raw frames $\hat{I}^{LR}_{j},j\in[1,...,M]$, where $M$ is the total number of raw frames in cryo-EM movies, have extremely low SNR and errors of SR reconstruction from LR images grow in proportion to the noise variance \cite{statsSr}, instead of using raw frames as LR inputs, we used averages of raw frames aligned to different reference frames:
\begin{equation}
    I^{LR}_{i} = F(\hat{I}^{LR}_{1,...,M},j),i=1,...,K 
\end{equation}
where $F$ aligns and averages raw frames $\hat{I}^{LR}_{1,...,M}$ with respect to the reference frame $j$. Reference frame $j$ is selected at random from all possible frames $M$. These generated LR frame averages $ I^{LR}_{i}$ have higher SNR, and since each $I^{LR}_{i}$ is aligned to a different reference frame, relative motions exist between these frame averages. Therefore, they are more suitable as LR inputs for SR reconstruction. \par 
The overall architecture of SR Net is based on HighResNet \cite{deudon2020highresnet}, which includes three main steps: encoding, fusion and decoding ({\bf Figure 2A}, left). The network learns to implicitly co-register multiple LR frames $I^{LR}_{i}$ and fuse them into a single SR view. Unlike HighResNet, which upscales input LR views during the decoding step using a deconvolution layer, we first upscale LR inputs to the desired SR output size before feeding into the encoder using bilinear interpolation. The network thus learns the residual between the interpolated LR and HR images. We also incorporate ShiftNet-Lanczos \cite{deudon2020highresnet} to account for pixel and sub-pixel shifts between generated SR and HR images, since these two images are not necessarily aligned ({\bf Figure 2A}, right). ShiftNet learns relative motions between two images and a Lanczos kernel is used to align generated SR to the HR image based on the output of ShiftNet. Motion compensated SR is then compared against the HR image. A detailed description of the architecture of ShiftNet is given in \cite{deudon2020highresnet}. \par
Training of the network is based on the ZSSR approach ({\bf Figure 2B}). As ground truth images do not exist for cryo-EM micrographs, a self-supervised algorithm that does not require knowledge of ground truth HR images is most suitable. For each set of LR images (we use $K=16$), a movie-specific SR Net is trained in the following way:
\begin{enumerate}
    \item Extract example patches of fixed size from input LR frames.
    \item Further downsample the extracted examples by a factor of $s$ (we use 2). These down-sampled examples now become temporary LRs. 
    \item From extracted examples, randomly select an example from a single frame. Note this selected example is not down-sampled and is treated as the temporary HR. 
    \item Feed temporary LR images obtained in step 2 into SR Net, a SR output is generated and compared with the temporary HR.
\end{enumerate}
The movie-specific SR Net is able to leverage the power of cross-scale internal recurrence of image specific information. To further enrich the training dataset, data augmentation is applied to the set of LR images to extract more pairs of HR-LR to train on, including mirror reflections in the vertical and horizontal directions as well as rotations. Once the network is trained, the series of LR frames $I^{LR}_{1,..K}$ are used as input to the network and the desired SR output $I^{SR}$ is constructed.\par 
The loss is computed as the mean squared error (MSE) between motion compensated generated SR and the actual HR images with total variation (TV) regularization. We use the ADAM optimizer. We start with a learning rate of 0.001 and adaptively decrease the learning rate based on the training procedure proposed in \cite{shocher2017zeroshot}. Training stops when the learning rate reaches $10^{-5}$, at around 200 epochs. The network is trained to learn upscaling by a factor of 2. At each iteration, a fixed crop size of $256 \times 256$ is used, while the 2X down-sampled versions have sizes of $128\times128$. This way, training time is independent of the input size. Finally, similar to \cite{shocher2017zeroshot}, we also adopt a geometric self-ensemble method (which generates 8 different outputs for the 8 rotations+flips of the test LR frames $I^{LR}_{1,..k}$, and then combines them). We take the median of these 8 outputs. We then further combine it with the back-projection technique \cite{backp1,backp2}. The final median image is also corrected by back-projection. Each set of LR images takes around 5 minutes to train for an upsampling factor of 2, the final SR image takes about 30 seconds to generate on an NVIDIA Tesla V100 GPU with 32GB of memory. 
\begin{figure}[t]
  \centering
  \includegraphics[width=13cm,height=15cm]{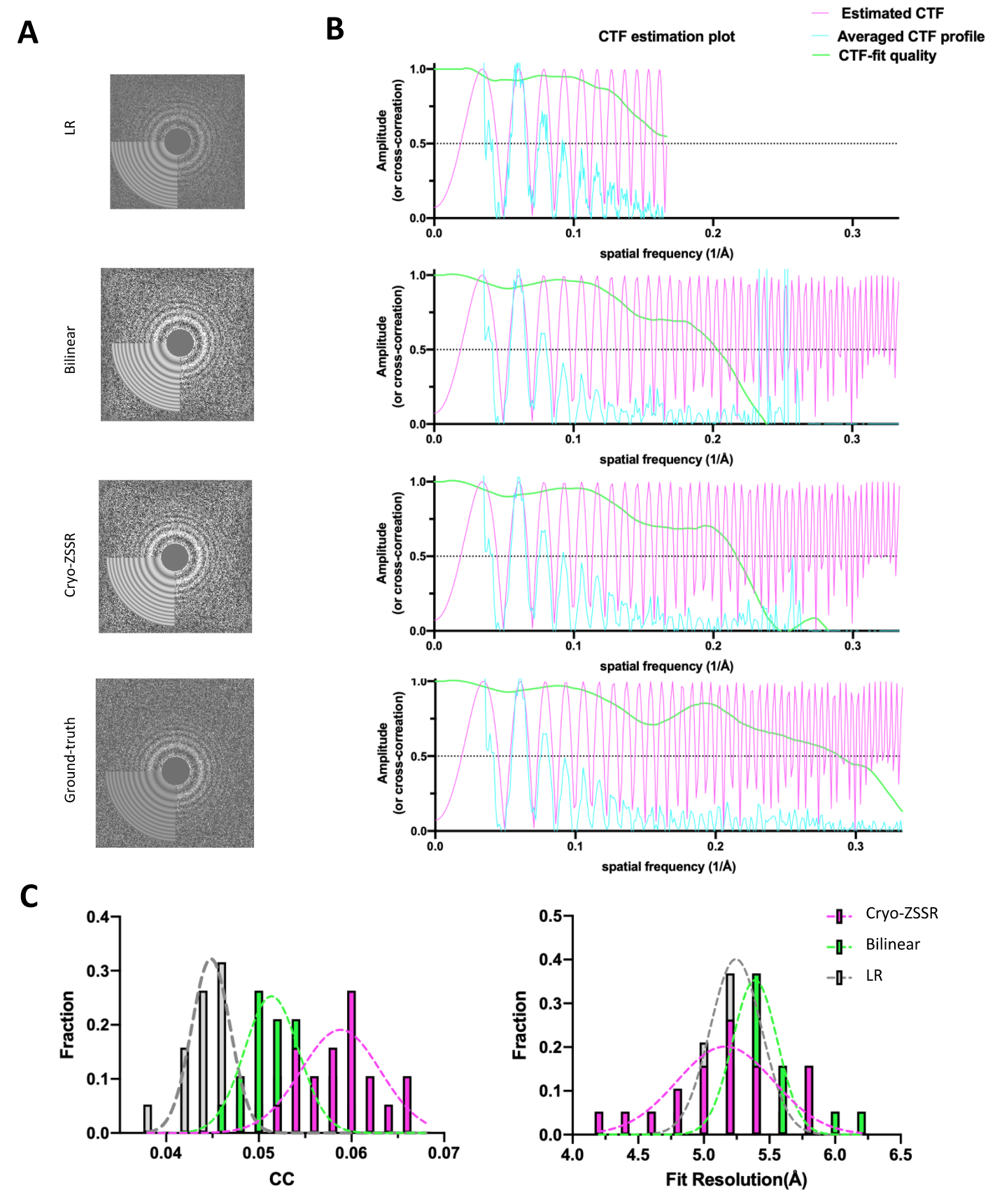}
  \caption{\underline{Cryo-ZSSR improves image quality metrics of individual micrographs.} To evaluate the performance of cryo-ZSSR at the micrograph level, we estimated the CTF of all 20 movies in the EMPIAR-10146 dataset before and after up-sampling. {\bf A-B}. 2D power spectrums ({\bf A}) and corresponding 1D radial profiles ({\bf B}) of original LR images (top row), up-sampled using bilinear interpolation (second row), up-sampled using cryo-ZSSR (third row), and ground truth (bottom row). Higher CTF fit quality represents better results. {\bf C}. Corresponding histograms of cross-correlation (CC) scores (left) and estimated fit resolution (right) showing the net improvement in image quality obtained by cryo-ZSSR (higher CC scores and lower fit resolutions represent better results).
}
\end{figure}
\begin{figure}[h!]
  \centering
  \includegraphics[width=13cm,height=15.5cm]{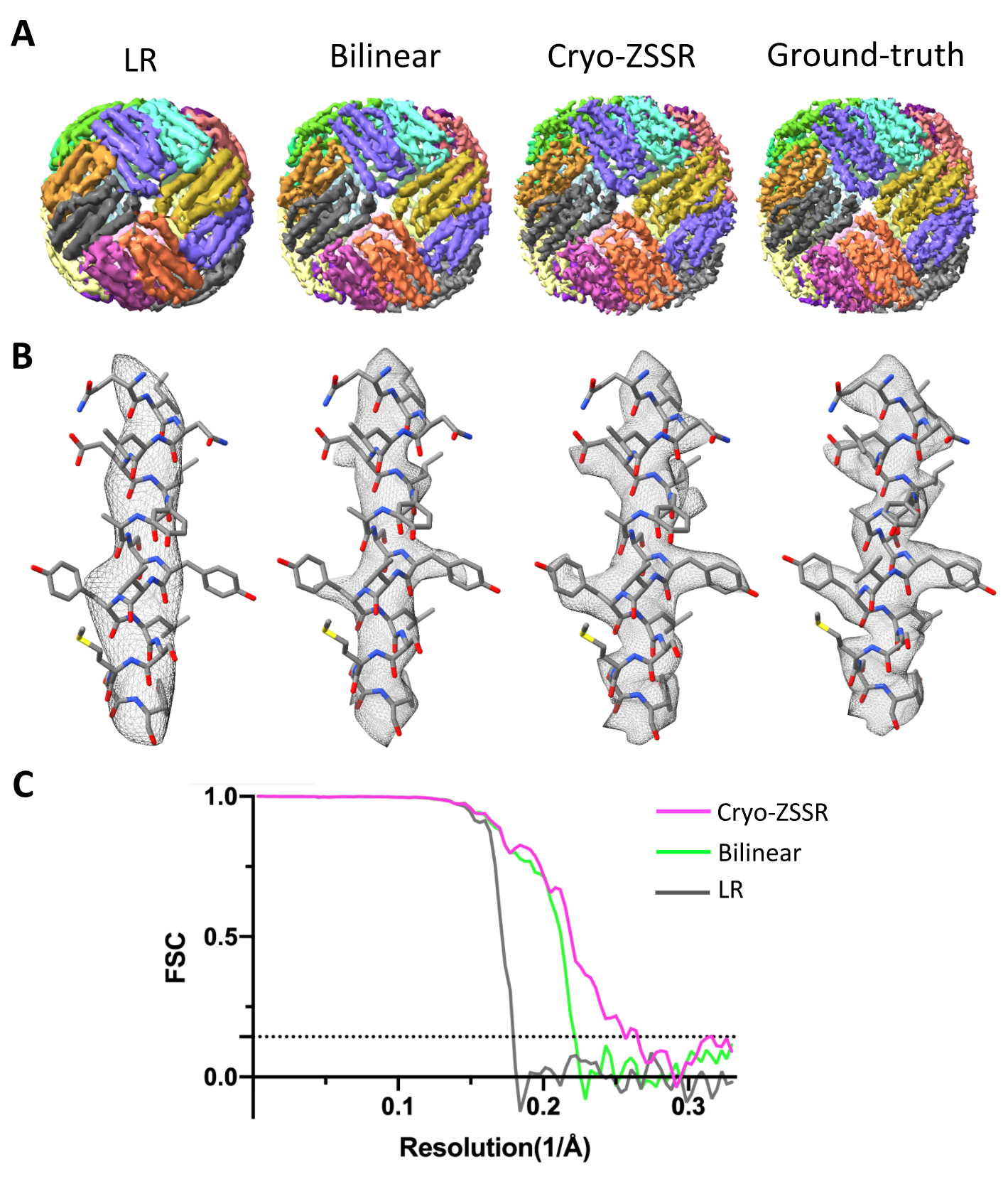}
  \caption{\underline{Cryo-ZSSR up-sampled images improve the resolution of apoferritin structure.} 1,200 particles were extracted from 19 movies (one low quality movie was discarded) and subjected to iterative 3D refinement for 8 iterations as implemented in the cisTEM package. {\bf A-B}. Overall structure of apoferritin (A) and zoomed-in view of an alpha helix with fitted atomic model (B), for maps obtained from the LR images (left column), up-sampled images using bilinear interpolation (second column), up-sampled images using cryo-ZSSR (third column), and ground truth images (last column). {\bf C}. Fourier Shell Correlation (FSC) curves for maps obtained using LR images (gray), up-sampled using bilinear interpolation (green), and up-sampled using cryo-ZSSR (magenta) against ground-truth reconstruction. Estimated resolutions are 5.9 {\AA}, 4.8 {\AA}, and 3.9 {\AA}, respectively based on the 0.143-cutoff (dotted line). Lower numbers represent better reconstruction quality.
}
\end{figure}
\section{Experiments and Validation on Single-Particle Cryo-EM Data}

To validate our approach, we used cryo-EM movies of apoferritin available from the Electron Microscopy Public Image Archive (EMPIAR) under accession code 10146 \cite{empiar}. Apoferritin is a commonly used test sample that has a molecular weight of 440 kDa and octahedral symmetry (O). This dataset consists of 20 movies with 50 frames each and $1240\times1200$ pixels in size. The physical pixel size is 1.5 {\AA} and the images were acquired using a beam energy of 300 kV and an exposure rate of 2 $e^-/${\AA}$^2$ (equivalent to a total dose of 100 $e^-$/{\AA}$^2$), {\bf Figure 1B-C}. 

Commonly used metrics to evaluate SR performance such as structural similarity index (SSIM), peak signal-to-noise ratio (PSNR) or even visual evaluation, although useful, do not directly reflect the quality of individual micrographs due to the very low SNRs. Instead, we turned to domain specific metrics used in cryo-EM, namely: 1) quality of CTF fit, 2) resolution of 3D reconstruction measured using the Fourier Shell Correlation (FSC) against the ground-truth, and 3) fit of atomic model coordinates into the cryo-EM density map.

The original movies (1.5 {\AA}/pixel) were cropped to dimensions of $1024\times 1024$ and subjected to the standard single-particle pipeline resulting in a 3.5 {\AA} resolution reconstruction from 1,200 particles that was used as ground-truth. We then binned the original movies by a factor of 2 using the IMOD program \cite{imod} (resulting in a pixel size of 3 {\AA}). down-sampled frames using uncropped frames had a size of $620\times600$ pixels and were aligned using MotionCorr2 \cite{motioncorr} and averaged. The frames were aligned to 16 different reference frames and the corresponding frame averages were generated. For convenience of implementation, we cropped patches of $512\times512$ from the 16 resulting frame averages and these patches were used as the input LR images $I^{LR}_{i}$. Movie-specific SR Nets were trained for each of the 20 movies. Once fully trained, LR images were upscaled by a factor of 2 through our framework so that $I^{SR}$ for each movie became $1024\times1024$ in size. The original movies (without downsampling) were never used or seen during any part of the training or testing steps. The 20 SR micrographs now replaced the down-sampled LR images and were used as inputs to the single-particle cryo-EM structure determination pipeline. The CTF of each SR micrograph was estimated using CTFFIND4 \cite{Rohou2015} and particles were extracted and subjected to iterative 3D refinement using the cisTEM package \cite{Grant2018}. This process was repeated for the LR images, and the SR micrographs up-sampled using bilinear interpolation and cryo-ZSSR.

\paragraph{Cryo-ZSSR improves the quality of individual micrographs.} To test the performance of our algorithm on individual micrographs, we estimated the CTF of each of the 20 micrographs using the three sets of images (LR, up-sampled using bilinear interpolation and cryo-ZSSR) and the ground truth. The CTF fit results for a representative micrograph are shown in {\bf Figure 3A-B} and indicate that the strength of the signal present in the cryo-ZSSR result is higher compared to the LR and bilinear interpolated images. We also quantified the overall improvement in image quality by tracking the CTF-fit values and estimated fit resolution for all 20 movies in the dataset, {\bf Figure 3C}. CTF-fit values reported by CTFFIND4 measure the quality of fit between the concentric Thon ring profile and the estimated CTF curve (higher scores represent better fits), while the estimated fit resolution gives an indication of how far the signal extends (lower numbers are better). Taken together, these metrics show that SR images reconstructed using our approach have higher CTF-fit scores and higher fit resolutions than both the LR and bilinear interpolated images, indicating that cryo-ZSSR can effectively extract high-resolution information present in the LR movie stacks. 

\paragraph{Cryo-ZSSR improves the resolution of 3D reconstructions.} We also evaluated the downstream effects of our SR interpolation algorithm by measuring the quality of the final 3D reconstructions. The three sets of movies (LR, up-sampled using bilinear interpolation and cryo-ZSSR) were used as input to the standard single-particle refinement pipeline implemented in cisTEM. 1,200 particles were selected and aligned against an external reference of apoferritin using iterative projection matching. We repeated this process using the three sets of images and the ground truth data, {\bf Figure 4A-B}. Consistent with the CTF estimation results, the features and resolution of the cryo-ZSSR map are better than the ones obtained using bilinear interpolation and the LR data, with estimated resolutions according to the 0.143-FSC criteria of 3.9 {\AA}, 4.8 {\AA} and 6.0 {\AA}, respectively ({\bf Figure 4C}). The resolution obtained using the ground truth images is 3.5 {\AA}.  Lower numbers indicate better reconstruction quality. The resolution obtained by cryo-ZSSR clearly surpasses the 6 {\AA} Nyquist limit imposed by the original physical pixel size of 3 {\AA}, and the reconstruction shows clear density for side chains, in agreement with the atomic model and corresponding structural features in the ground-truth map.

\section{Conclusions}

We present a neural network framework to up-sample low-SNR single-particle cryo-EM movies using a multiple-image super-resolution algorithm based on the self-supervised deep internal learning approach. We achieve this without the need for ground-truth or prior training using HR images. Application of this technique to a LR dataset of apoferritin sampled at 3 {\AA}/pixel resulted in a three-dimensional reconstruction at 3.9 {\AA} resolution where side chains could be visualized at a similar level of detail seen in the ground-truth map. This result suggests that our SR strategy may be used in conjunction with lower magnification imaging to accelerate data collection without sacrificing image quality. Our algorithm does not presently account for the resolution-lowering effects caused by radiation damage affecting cryo-EM samples. Future efforts to incorporate this feature into our framework could result in further improvements in resolution. While our end goal is to obtain SR images using the ground-truth data as LR images, our present implementation is not yet optimized to handle images exceeding 2k pixels in size. We are also currently investigating whether the application of cryo-ZSSR can be extended to work on full-size cryo-EM datasets consisting of several thousand micrographs. In addition, we will investigate how protein symmetry can affect the performance of the algorithm, as symmetry plays an important role in internal statistics recurrence. Finally, our results show that cryo-ZSSR is an effective strategy to recover high-resolution information contained in low-SNR cryo-EM movies and can be used to produce high-resolution 3D structures from LR micrographs. 

\section*{Acknowledgments}

This study utilized the computational resources offered by Duke Research Computing (\texttt{http://rc.duke.edu}). We thank Mark DeLong, Charley Kneifel, Mike Newton, Victor Orlikowski, Tom Milledge, and David Lane from the Duke Office of Information Technology and Research Computing for providing assistance with the computing environment.

\bibliography{bibliography}

\end{document}